\documentclass[conference]{IEEEtran}
\IEEEoverridecommandlockouts
\usepackage{cite}
\usepackage{amsmath,amssymb,amsfonts}
\usepackage{algorithmic}
\usepackage{graphicx}
\usepackage{textcomp}
\usepackage{xcolor}
\usepackage{hyperref}
\usepackage{CJKutf8}

\def\BibTeX{{\rm B\kern-.05em{\sc i\kern-.025em b}\kern-.08em
    T\kern-.1667em\lower.7ex\hbox{E}\kern-.125emX}}
\begin{document}
\begin{CJK}{UTF8}{gbsn}
\title{Rotatable Antenna Enabled Wireless Communication System with Visual Recognition: A Prototype Implementation
\vspace{-0.35cm}}
\author{
\IEEEauthorblockN{Liang Dai\IEEEauthorrefmark{1}, Beixiong Zheng\IEEEauthorrefmark{1}, Yanhua Tan\IEEEauthorrefmark{1}, Lipeng Zhu\IEEEauthorrefmark{4}, Fangjiong Chen\IEEEauthorrefmark{2}, Rui Zhang\IEEEauthorrefmark{3}\IEEEauthorrefmark{4}}
\IEEEauthorblockA{\IEEEauthorrefmark{1}School of Microelectronics, South China University of Technology, Guangzhou 511442, China}
\IEEEauthorblockA{\IEEEauthorrefmark{2}School of Electronic and Information Engineering, South China University of Technology, Guangzhou 510641, China}
\IEEEauthorblockA{\IEEEauthorrefmark{3}School of Science and Engineering, The Chinese University of Hong Kong, Shenzhen 518172, China}
\IEEEauthorblockA{\IEEEauthorrefmark{4}Department of Electrical and Computer Engineering, National University of Singapore, Singapore 117583, Singapore
}
Email: 202321061996@mail.scut.edu.cn; bxzheng@scut.edu.cn; tanyanhua06@163.com; zhulp@nus.edu.sg;\\ eefjchen@scut.edu.cn; rzhang@cuhk.edu.cn
\vspace{-0.55cm}
}

\maketitle

\begin{abstract}
Rotatable antenna (RA) is an emerging technology that has great potential to exploit additional spatial degrees of freedom (DoFs) by flexibly altering the three-dimensional (3D) orientation/boresight of each antenna. In this demonstration, we present a prototype of the RA-enabled wireless communication system with a visual recognition module to evaluate the performance gains provided by the RA in practical environments. In particular, a mechanically-driven RA is developed by integrating a digital servo motor, a directional antenna, and a microcontroller, which enables the dynamic adjustment of the RA orientation. Moreover, the orientation adjustment of  the RA is guided by the user's direction information provided by the visual recognition module, thereby significantly enhancing system response speed and self-orientation accuracy. The experimental results demonstrate that the RA-enabled communication system achieves significant improvement in communication coverage performance compared to the conventional fixed antenna system.

\end{abstract}

\begin{IEEEkeywords}
Rotatable antenna, visual recognition, 3D orientation.
\end{IEEEkeywords}

\section{Introduction}
The forthcoming wireless communication network is envisioned to support intelligent connectivity for more devices and users, demanding significant enhancements in transmission rate, service range, connectivity scale, and reliability. Currently, large-scale multiple-input multiple-output (MIMO) has become a key physical layer technology to dramatically enhance the transmission rate and reliability of wireless networks. Although large-scale MIMO offers substantial array beamforming and spatial multiplexing gains, it poses several challenges, including increased array size, higher energy consumption, and escalating hardware cost. In particular, as the number of antennas increases while their deployed area is fixed, mutual coupling and interference between antennas intensify, necessitating more stringent requirements for hardware design and calibration. Furthermore, increasing the number of antennas cannot fully exploit the spatial degrees of freedom (DoFs) due to the fixed position and orientation of each antenna. To overcome this limitation, rotatable antenna (RA)\cite{4}\cite{3} has been proposed to improve the performance of wireless communications without increasing the number of antenna. Specifically, RA can enhance three-dimensional (3D) coverage by flexibly adjusting its 3D orientation while maintaining its 3D position. Thus, it can be implemented with lower hardware cost and time/energy overhead as compared to both antenna position and rotation adjustments. In particular, RA can also be considered as a simplified implementation of the more general six-dimensional movable antenna (6DMA) \cite{8} with antenna rotation only. Despite its promising benefits, the development of a practical prototype of RA-enabled communication system has not been pursued yet, which is of great significance for its performance validation and practical deployment in realistic environments.

In this paper, a prototype of RA-enabled wireless communication system is presented, including its overall architecture, constituting modules, and operation algorithms. Specifically, the RA system prototype integrates a digital servo motor, a directional antenna, and a microcontroller. Additionally, a camera, a personal computer (PC) and a universal software radio peripheral (USRP) are employed to process user angular information and radio frequency (RF) signal in this prototype. Notably, the antenna orientation is dynamically controlled in real time by the servo, based on the precise user direction information provided by a separate vision recognition module. This design significantly enhances the RA system response speed and self-orientation accuracy.

\section{Implementation}
	\vspace{-0.3cm}
\begin{figure}[h]
	\centering
	\includegraphics[width=2.5in]{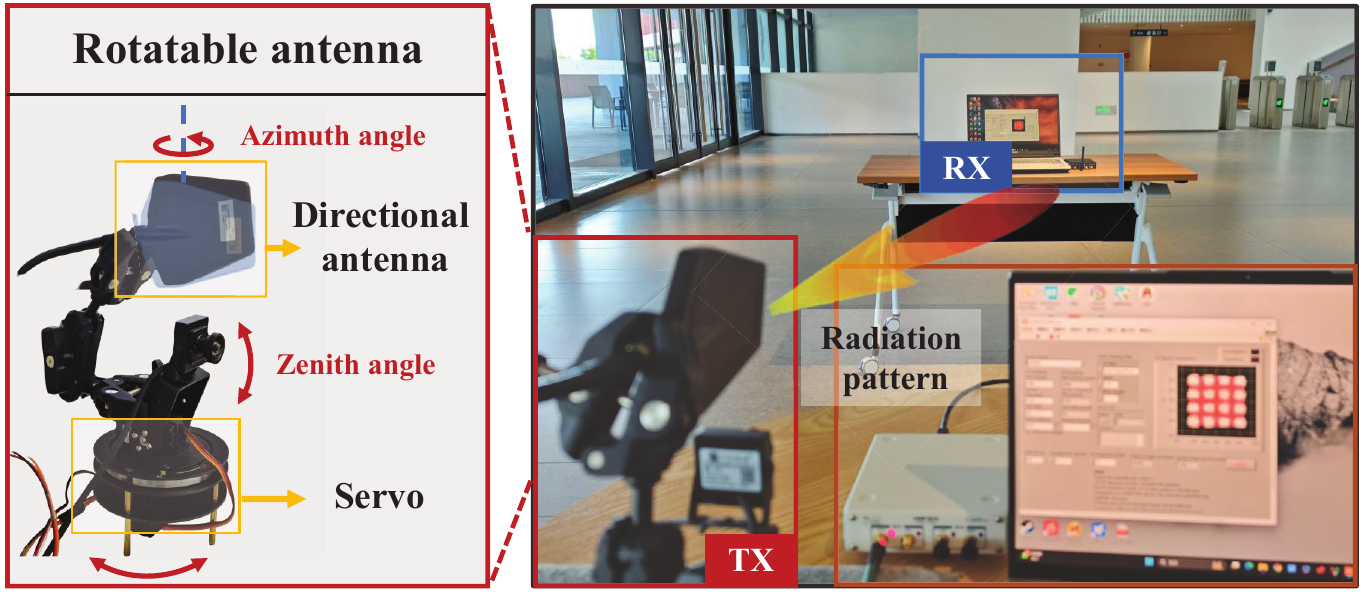}
	\setlength{\abovecaptionskip}{-4pt}
	\vspace{0.2cm}
	\caption{Mechanically-driven rotatable antenna communication prototype}
	\label{fig1}
	\vspace{-0.2cm}
\end{figure}
Fig. \ref{fig1} illustrates the proposed RA-enabled wireless communication system. The transmitter in the prototype comprises a USRP for radio signal processing, a PC for control and computation, a vision recognition module for user identification and tracking, and an RA for directional signal transmission. The user receiver consists of an isotropic antenna for signal reception, another USRP for signal processing, and a PC for signal analysis and visualization. The two main components of the prototype are elaborated as follows:

\begin{itemize}
\item{\bf{Rotatable antenna (RA)}}: It consists of a directional antenna with a 10 dBi gain and a 60° beamwidth, a two-dimensional digital servo responsible for precisely controlling the directional antenna’s zenith angle and azimuth angle, and a microcontroller that manages the servo's rotation. The control circuit within the servo converts the output pulse signal from the microcontroller into a current signal, leveraging electromagnetic induction to drive the integrated motor to rotate, thereby reorienting the antenna toward the desired direction. Meanwhile, the current direction of the antenna is continuously monitored by an angle sensor and compared to the desired angle by the control circuit, forming a closed-loop control system to achieve the precise angle control. 
\item{\bf{Vision recognition module}}: It comprises a camera and a PC. The camera conveys red-green-blue (RGB) images to the PC, which processes the visual measurement data and subsequently transmits the results to the microcontroller. The PC employs an efficient object detection network named “you only look once (YOLO)”\cite{6} that utilizes convolutional neural networks to extract multi-scale features from input images to detect the user. Subsequently, a tracking-by-detection algorithm named “DeepSORT”\cite{7} is employed to ensure the accurate tracking direction variations for the moving user. The direction of the user can be ascertained from the vision recognition module.
\end{itemize}\par
\vspace{-0.3cm}
\begin{figure}[htbp]
	\centering
	\includegraphics[width=2.8in]{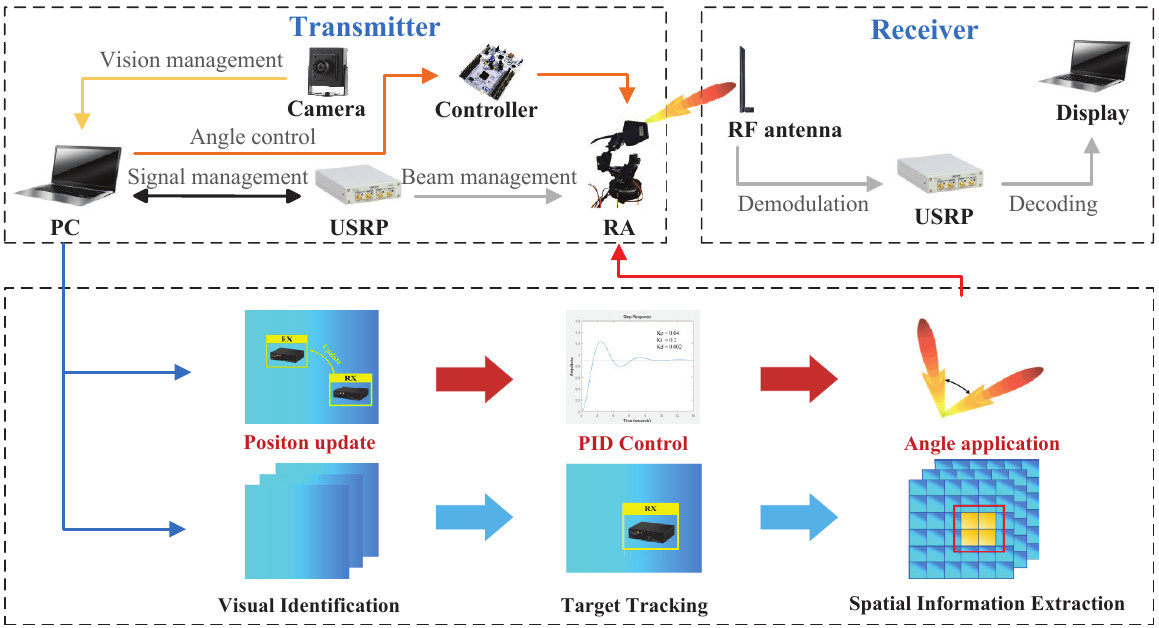}
	\setlength{\abovecaptionskip}{-4pt}
	\vspace{0.2cm}
	\caption{System structure of RA-enabled communication system prototype}
	\label{fig2}
	\vspace{-0.2cm}
\end{figure}

 The implementation details of the RA-enabled communication system are shown in Fig.~\ref{fig2}. The system leverages vision recognition module for identifying and tracking the user, effectively transforming the angle estimation problem in RF into an object detection problem on images. When no user is detected within the camera’s visual range, the PC periodically activates the camera to perform environmental scanning, employing deep learning models to detect and identify any potential user. Once the user comes into the camera’s visual range, the vision recognition module locks onto the user and processes its directional information. Subsequently, the PC adjusts the RA orientation by modifying the output pulse width of the controller, ensuring precise alignment of the RA orientation with the user direction for enhancing communication performance. When the user moves to another direction within the camera's visual range, a user tracking algorithm is triggered to ascertain the real-time direction of the user. Then, a proportional-integral-derivative (PID) steering algorithm based on the user's updated direction is employed to ensure precise rotation of the RA to align its boresight direction with that of the user over time.

\section{Experimental Results}
\vspace{-0.5cm}
\begin{figure}[!h]
	\centering
	\includegraphics[width=2.4in]{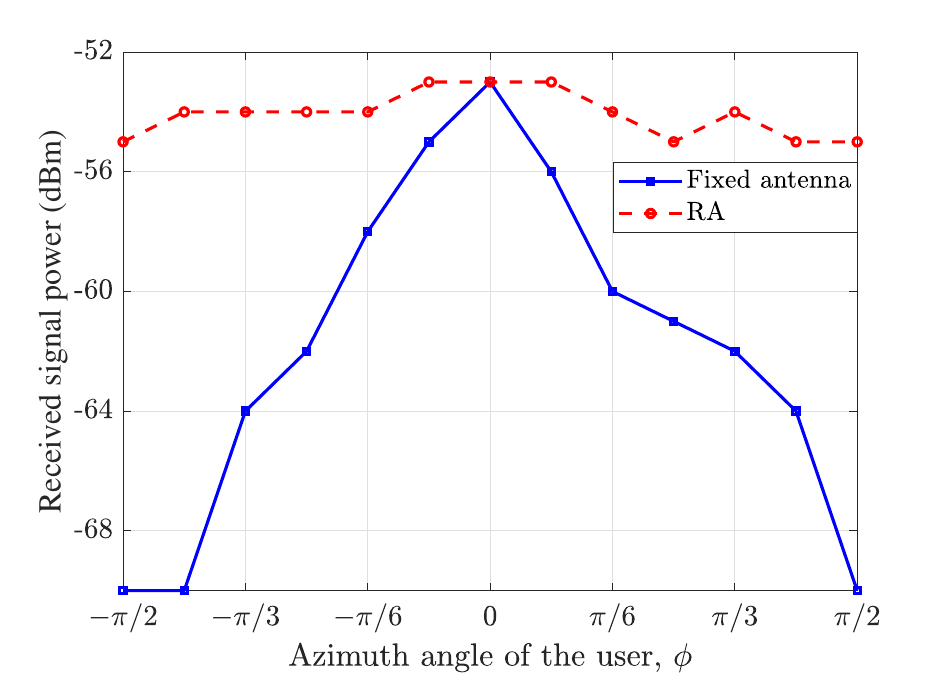}
	\setlength{\abovecaptionskip}{-4pt}
	\vspace{-0.05cm}
	\caption{The received signal power versus the azimuth angle of the user}
	\label{fig3}
	\vspace{-0.2cm}
\end{figure}
\vspace{-0.1cm}
In the experiment, the RA operates at a carrier frequency of 5.8 GHz and adopts the 16-quadrature amplitude modulation (QAM). The transmit power and transmission rate are set to 10 dBm and 2 Mbps, respectively. Fig. \ref{fig3} shows the the user's received power versus its azimuth angle, while the zenith angle of the user is fixed at $\theta = 0^{\circ}$. As the user's azimuth angle varies from $-\frac{\pi}{2}$ to $\frac{\pi}{2}$, the RA dynamically adjusts its orientation/boresight to align with the user's direction in real time, ensuring that the user receives a stable signal power over time. In contrast, for a fixed antenna system (i.e., the antenna orientation is fixed and cannot be adjusted over time), the user's movement causes a discrepancy between the orientation/boresight of the antenna and the user direction, thus leading to a significant degradation in the received signal power. The experimental results demonstrate the effectiveness of the proposed RA with camera-aided antenna orientation/boresight adjustment to enhance communication coverage performance. A supplementary video is available at  \url{https://youtu.be/r4zzDOmQ2ZI} or \url{http://www.iqiyi.com/v_23qcq5oyjr0.html}.

\section{Conclusion}
In this demonstration, we developed a prototype of the RA-enabled wireless communication system for the first time, with precise user detection and automatic rotation steering. The RA system was deployed and tested in realistic environment with a moving communication user. Experimental results verified the RA's capability to significantly enhance received signal power as compared to conventional fixed-antenna system. 

\bibliography{ref}

\begin{thebibliography}{1}
\providecommand{\url}[1]{#1}
\csname url@samestyle\endcsname
\providecommand{\newblock}{\relax}
\providecommand{\bibinfo}[2]{#2}
\providecommand{\BIBentrySTDinterwordspacing}{\spaceskip=0pt\relax}
\providecommand{\BIBentryALTinterwordstretchfactor}{4}
\providecommand{\BIBentryALTinterwordspacing}{\spaceskip=\fontdimen2\font plus
\BIBentryALTinterwordstretchfactor\fontdimen3\font minus
  \fontdimen4\font\relax}
\providecommand{\BIBforeignlanguage}[2]{{%
\expandafter\ifx\csname l@#1\endcsname\relax
\typeout{** WARNING: IEEEtran.bst: No hyphenation pattern has been}%
\typeout{** loaded for the language `#1'. Using the pattern for}%
\typeout{** the default language instead.}%
\else
\language=\csname l@#1\endcsname
\fi
#2}}
\providecommand{\BIBdecl}{\relax}
\BIBdecl

\bibitem{4}
Q.~{Wu}, B.~{Zheng}, T.~{Ma}, and R.~{Zhang}, ``{Modeling and optimization for
  rotatable antenna enabled wireless communication},'' \emph{arXiv preprint
  arXiv:2411.08411}, 2024.

\bibitem{3}
B.~{Zheng}, Q.~{Wu}, and R.~{Zhang}, ``{Rotatable antenna enabled wireless
  communication: modeling and optimization},'' \emph{arXiv preprint
  arXiv:2501.02595}, 2025.

\bibitem{8}
X.~{Shao}, Q，{Jiang}, and R.~{Zhang}, ``{6D movable antenna based on user
  distribution: modeling and optimization},'' \emph{IEEE Trans. Wireless
  Commun.}, vol.~24, no.~1, pp. 355--370, Jan. 2025.

\bibitem{6}
J.~{Redmon}, S.~{Divvala}, R.~{Girshick}, and A.~{Farhadi}, ``{You only look
  once: unified, real-time object detection},'' in \emph{Proc. IEEE Conf.
  Comput. Vis. Pattern Recog}, Jun. 2016, pp. 779--788.

\bibitem{7}
N.~{Wojke}, A.~{Bewley}, and D.~{Paulus}, ``{Simple online and realtime
  tracking with a deep association metric},'' in \emph{Proc. Int. Conf. Image
  Process}, Sept. 2017, pp. 3645--3649.

\end{thebibliography}
\end{CJK}
\end{document}